\RequirePackage{fix-cm}
\documentclass[natbib,smallextended]{svjour3} 

\smartqed  

\usepackage{graphicx}
\usepackage{bbold}
\usepackage{hyperref}
\usepackage{amsmath,amssymb}
\hypersetup{
    colorlinks,%
    citecolor=blue,%
    linkcolor=blue,%
    urlcolor=blue
}
\usepackage{xcolor}

\usepackage{ulem}
\usepackage{aps-bibstyle}  

\newcommand{\xraychap}{X-ray Chapter}

\newcommand{\optchap}{Optical Chapter}
\newcommand{\radiochap}{Radio Chapter}

\newcommand{\impostchap}{Imposters Chapter}

\newcommand{\disrupchap}{Disruption Chapter}
\newcommand{\flowchap}{Formation of the Accretion Flow Chapter} 

\newcommand{\simchap}{Simulation Methods Chapter}
\newcommand{\emischap}{Emission Mechanisms Chapter}

 \journalname{ISSI Book on TDEs}

\usepackage{ref_macros} 

\begin{document}

\title{The Physics of Accretion Discs, Winds And Jets in Tidal Disruption Events}


\titlerunning{TDE Accretion Physics}        

\author{Jane Lixin Dai \and
        Giuseppe Lodato \and
        Roseanne Cheng 
}

\authorrunning{J. Dai et al.} 

\institute{J. L. Dai \at
              Department of Physics\\ 
              The University of Hong Kong\\ Pokfulam Road, Hong Kong (HK) \\  \email{lixindai@hku.hk} 
            \and
            G. Lodato \at
              Universit\`a degli Studi di Milano \\
              Dpartimento di Fisica\\
              Via Celoria 16, Milano (Italy)\\
              \email{giuseppe.lodato@unimi.it}   
            \and
            R.~M.~Cheng \at
              Theoretical Division (T-3)\\
              Los Alamos National Laboratory\\
              P.O. Box 1663\\
              Los Alamos, NM 87545 (USA)\\
              \email{rmcheng@lanl.gov}
}

\date{Received: date / Accepted: date}

\maketitle

\begin{abstract}
Accretion onto black holes is an efficient mechanism in converting the gas mass-energy into energetic outputs as radiation, wind and jet.
Tidal disruption events, in which stars are tidally torn apart and then accreted onto supermassive black holes, offer unique opportunities of studying the
accretion physics as well as the wind and jet launching physics across different
accretion regimes. In this review, we systematically describe and discuss the models that have been developed to study the accretion flows and jets in tidal disruption events. A good knowledge of these physics is not only needed for understanding the  emissions of the observed events, but also crucial for probing the general relativistic space-time around black holes and the demographics of supermassive black holes via tidal disruption events.


\end{abstract}

\newpage
\section{Introduction}
A tidal disruption event (TDE) happens when a star passes by a supermassive black hole (SMBH) too close and is therefore torn apart by the tidal force. The distance at which the SMBH tidal force equals the stellar self-gravity is called the tidal radius $R_t$ \citep{Hills75}, which depends on the mass $M_\star$ and radius $R_\star$ of the star as well as the mass $M_{\rm BH}$ of the SMBH: $R_t \approx R_\star~(M_{\rm BH}/{M_\star})^{1/3}$. For main-sequence stars approaching SMBHs with $M_{\rm BH}={\rm few}\times(10^6-10^7) M_\odot$, $R_t$ is only ${\rm few}\times (1-10) R_g$ with $R_g\equiv G M_{\rm BH}/c^2$ being the gravitational radius of the black hole. Therefore, general relativity (GR) naturally plays an important role in the processes happening in TDEs. 

The basic theory of TDE physics has been developed since the 1970s \citep[e.g.,][]{Hills75,Rees88,Phinney89}. Here we give a brief description of the general picture. Following the disruption of a star approaching the black hole along a marginally bound orbit, roughly half of the stellar debris remains bound and travels along highly elliptical, ballistic orbits, while the other half escapes along hyperbolic orbits. The spread of the orbital energy determines the mass rate that bound debris returns to the pericenter as a function of time, which is called the fallback rate. Shortly after passing through the pericenter, the bound debris assembles a transient accretion disc around the SMBH through shocks and shear viscosity reducing the orbital energy of the debris. Then with the aid of magnetic fields, viscous processes transport the gas angular momentum outwards, allowing gas to accrete onto the black hole and bright flares to be produced. Under optimal conditions, winds and relativistic jets can also be launched from the transient discs formed in TDEs.

In recent years our understanding of TDE physics has been greatly pushed forward by numerical simulations. These simulations usually focus on specific phases in TDEs (e.g., the disruption of the star, the evolution of the debris stream, the formation of the disc, the accretion process and jet production, etc.) due to the large computational resources required to study the whole process. These studies have basically confirmed the disruption process described in the general picture, which is discussed in the \disrupchap~of this book.
However, up to now no self-consistent simulations have been done on the disc formation process for typical TDE parameters, leading to uncertainties on whether the debris can quickly dissipate orbital energy to form discs as described in the general picture. We refer the readers to the \flowchap~for details. In this Chapter dedicated to the TDE accretion process, we will focus on the phase after stable discs have been formed and review the theoretical efforts that have been made to investigate the physics of TDE accretion discs, winds and jets. 

\begin{figure}
\begin{center}
\includegraphics[width=0.8\textwidth]{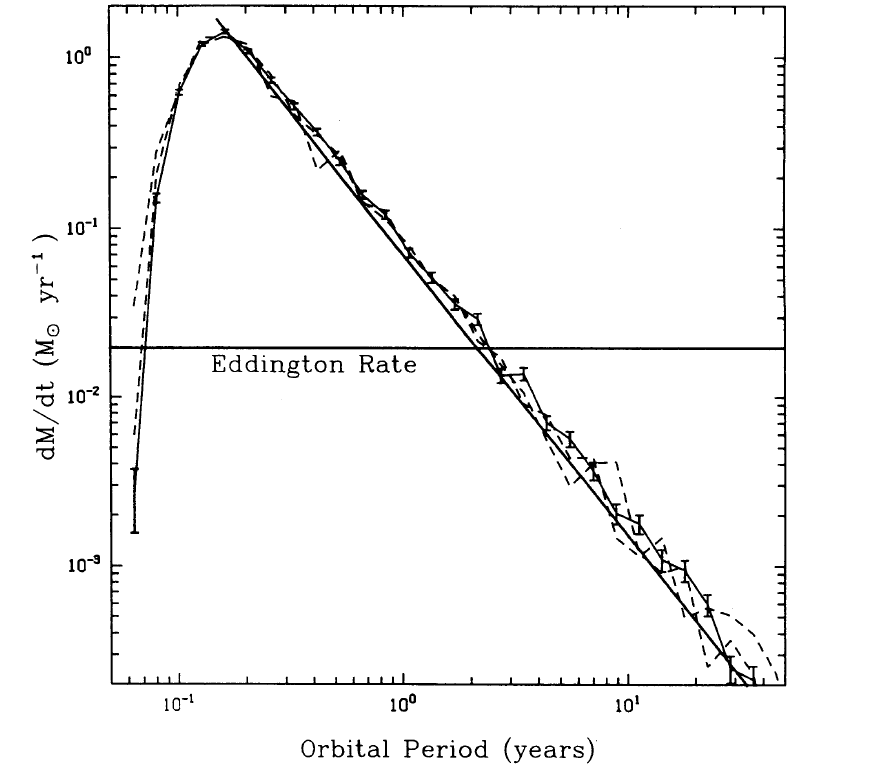}
\caption{The debris fallback rate as a function of time. The Eddington accretion rate shown is for a SMBH with $M_{\rm BH}= 10^6M_\odot$ and radiative efficiency $\eta=0.1$. As shown the fallback rate can be one to two orders of magnitude above the Eddington accretion rate of the black hole when $M_{\rm BH} < {\rm few}\times10^7 M_\odot$. After the peak the fallback rate declines following a power-law function of time: $\dot{M}_{\rm fb}\propto t^{-5/3}$, sweeping through different regimes of accretion. Figure is from \citet{Evans89}.}
\label{fig:fallback}
\end{center}
\end{figure}

A solid understanding of the accretion process builds foundations for understanding TDE emissions, and therefore is key for probing SMBH demographics from TDE observables. Also, TDEs give us ideal laboratories for studying black hole accretion physics in different accretion regimes. In a standard TDE (where a main-sequence star is totally disrupted by a SMBH along a parabolic orbit), the debris fallback rate that feeds the disc is expected to first quickly reach the super-Eddington regime, and then declines to the sub-Eddington regime over a few months to years (Fig. \ref{fig:fallback}).

The Eddington luminosity is the theoretical maximum luminosity that a star or an accretion disc can have when there is a balance between the outward radiation force and the inward gravitational force. For the inner part of accretion discs, all of the hydrogen atoms are ionized since the temperature is higher than $10^5$K \citep{Shakura73, Frank02}. The Eddington luminosity for a black hole is therefore defined as:
\begin{equation}
    L_{\rm Edd}=\frac{4\pi c  G  M_{\rm BH} }{\kappa_{\rm es}}\approx 1.26\times10^{44}\left(\frac{M_{\rm BH}}{10^6M_{\odot}}\right) \mbox{erg}~\mbox{s}^{-1}.
\end{equation}
Here $\kappa_{\rm es}$ is the Thomson scattering opacity. For hydrogen $\kappa_{\rm es} ={\sigma_{\rm T}/ m_{\rm p}} = 0.4~\rm{cm^2 g^{-1}}$, where $m_{\rm p}$ is the proton mass and $\sigma_{\rm T}$ is the Thomson cross-section. 
The corresponding Eddington accretion rate is:
\begin{equation}
    \dot{M}_{\rm Edd}=\frac{L_{\rm Edd}}{\eta c^2}\approx 0.022\left(\frac{\eta}{0.1}\right)^{-1}\left(\frac{M}{10^6M_{\odot}}\right)M_{\odot}~\mbox{yr}^{-1} ,
\end{equation}
where $\eta\equiv~L/(\dot{M}c^2)$ is the accretion radiative efficiency. In literature $\eta$ is usually given a nominal value of $10\%$, although its value can vary drastically as described next.

The structure of an accretion disc is primarily determined by its Eddington ratio, which is how the accretion rate compares to the Eddington accretion rate. (1) For very high, super-Eddington rates ($\dot{M}/\dot{M}_{\rm Edd} \gtrsim 1$), the disc is expected to be geometrically and optically thick. The photon diffusion time is very long and the disc is dominated by advection (the so-called slim disc solutions by \citealt{Abramowicz88}) resulting in low radiative efficiency; (2) for moderate accretion rates, $0.01\lesssim \dot{M}/\dot{M}_{\rm Edd}\lesssim1$ (or $0.01\lesssim L/L_{\rm Edd}\lesssim 0.3$), the disc, well described by the standard \citet{Shakura73} solutions (or a GR prescription by \citealt{Novikov73}), is optically thick, geometrically thin, and radiatively efficient; finally, (3) for very low accretion rates ($\dot{M}/\dot{M}_{\rm Edd} < 0.01$), the disc transitions again to an advection-dominated accretion flow (ADAF) or a radiatively inefficient accretion flow (RIAF) \citep{Ichimaru77,Rees82, Narayan94,Blandford99,Blandford04} with low radiative efficiency again. Although the Eddington ratio likely plays the most important role in determining the morphology of an accretion disc, the properties of the disc are also affected other physical parameters. For example, the black hole mass, together with the accretion rate, determine the disc temperature and various opacities governing disc dynamics and radiative processes. Also, in the thin disc regime, the inner edge of the disc is set by the innermost stable circular orbit (ISCO) which has a strong dependence on the black hole spin. As a result, $\eta$ varies from $6\%$ to $42 \%$ as the black hole spin increases from 0 to 1. 

\begin{figure}
\begin{center}
\includegraphics[width=0.8\textwidth]{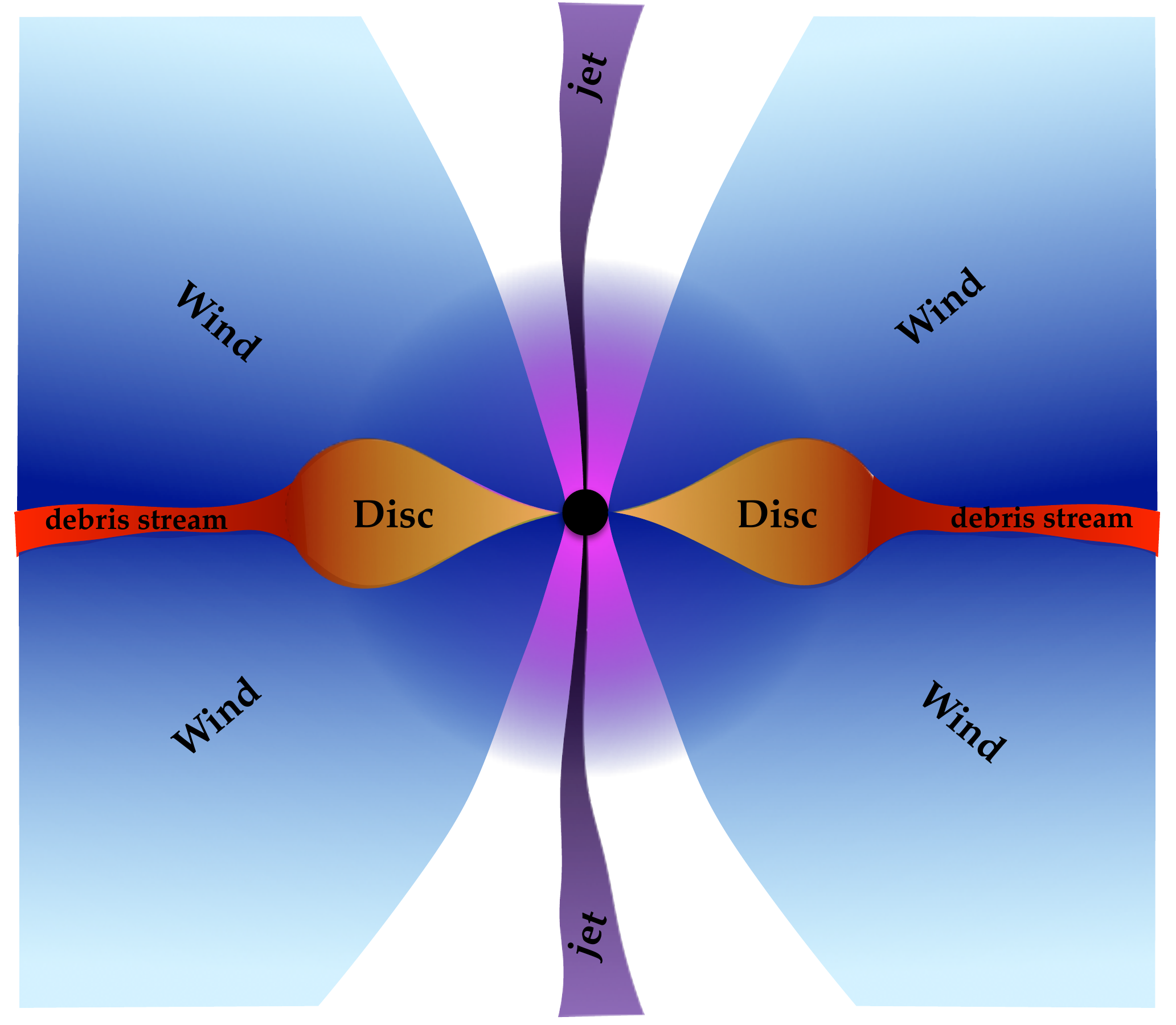}
\caption{A cartoon showing the geometry of a circular and aligned geometrically-thick accretion disc formed in the super-Eddington
phase of a TDE, as well as the wind and jet launched from the disc. It is expected that a super-Eddington disc can generally launch optically-thick wind with large solid angle coverage. A strong relativistic jet, on the other hand, can only be produced when the black hole is spinning fast and has a large amount of ordered magnetic flux in its vicinity \citep{Blandford77}. Figure is adapted from \citet{Dai18}.}
\label{fig:discsuperedd}
\end{center}
\end{figure}

\citet{Ulmer99} is the first to point out that different regimes of accretion should naturally happen throughout the course of a TDE, so the disc geometry should change from thick to thin as the accretion level drops from super-Eddington to sub-Eddington in a few months to years. Analytical models that encompass both the slim disc case and the standard thin disc case have been first constructed by \citet{Strubbe09}. We will review various disc models and simulations along this line in Section \ref{sec:superedd} (early, super-Eddington phase) and Section \ref{sec:thindisc} (late time). New disc models or angular momentum transport mechanisms different from the conventional picture will be reviewed in Section \ref{sec:eccentric} and Section \ref{sec:newLtransport}. 

Relativistic jets and winds have been observed in several TDEs, which cannot been explained using the standard thin disc theory. In Section \ref{sec:wind} and Section \ref{sec:tdejet}, we will review works which discuss how winds and jets can be launched when TDE disc is in the super-Eddington accretion regime. A cartoon illustrating the disc-wind-jet geometry during this phase is illustrated in Fig.~\ref{fig:discsuperedd}.

Most TDE literature have so far focused on modeling discs which have angular momentum aligned with the spin axis of the black hole. However, misaligned TDE discs likely can form since it is generally expected that stars are scattered into the loss cone from random orientations. We will discuss works that have been carried out to understand such discs and their jets in Section \ref{sec:discalign}.

Lastly, a summary will be drawn in Section \ref{sec:summary}, 
where we will also raise several key open questions related to TDE accretion discs, winds and jets for future studies.

\section{Models of TDE Accretion Flow}
\label{sec:accretionmodel}

In this section, we will summarize various analytical models and numerical simulations that have been carried out to study the accretion flow formed in TDEs. 
While we put the focus on describing the structures of the accretion inflow/outflow and the accretion process, we will also briefly mention the properties of the emissions produced in these models. We refer the readers to the \emischap~in this book for more details of the radiative processes in TDE accretion flow.

\subsection{TDE discs with super-Eddington debris fallback}
\label{sec:superedd}
During the first few weeks to years in a TDE, the debris supply rate is high and can exceed the Eddington accretion rate of the black hole if the black hole mass $M_{\mathrm{BH}}\lesssim {\mathrm{few}}\times 10^{7}~M_{\odot }$. It is naturally expected that most TDEs are detected in this phase as the flares are luminous. Many theoretical studies have therefore focused on studying the properties of the TDE discs in this phase.
Most of the work are semi-analytical, in which the TDE accretion flow is modeled as a steady, quasi-spherical disc \citep{Loeb97,Ulmer99,Coughlin14,Roth16, Wu18} or a disc surrounded by a quasi-spherical wind
\citep{Strubbe09,Lodato:2011a,Shen:2014b,MillerCole15, Metzger16}. Recently 3D numerical simulations of super-Eddington accretion flow have also been carried out to study TDE discs, in which the disc winds (and jets) are self-consistently produced \citep{Dai18, Curd19}.

From the observational side, the unique properties of TDE emissions also lend strong support to the notion that TDE discs should be different from the discs of local AGNs, which are believed to be sub-Eddington in most cases. Some of the most relevant observational features are: 
1) Many TDE flares are strong in UV/optical wavebands but weak in X-rays, while some TDEs are strong in X-rays but not in UV/optical bands. 
2) For the optically-strong TDEs, their flares are usually bluer than AGN optical emissions. There are three spectroscopic classes discovered: TDEs that produce broad H Balmer emission lines, He emission lines, or Bowen fluorescent NIII/OIII emission lines \citep{Gezari12, Arcavi14, Leloudas19, vanVelzen20}. 
3) For the non-jetted X-ray strong TDEs, their emissions are dominated by thermal components and do not show strong coronae which are typically seen in AGNs \citep{Komossa15review, Auchettl17}. 4) In some TDEs, fast winds have been reported, which will be discussed more in Section \ref{sec:wind}. More details on how TDE emissions are distinctive from AGN emissions can be found in the \impostchap.
The disc models to be discussed in this section all employ extreme debris supply or accretion level to explain these observed TDE features.

\subsubsection{Quasi-spherical models of steady discs or envelopes}
\label{sec:steadymodel}

\begin{figure}
\begin{center}
\includegraphics[width=0.8\textwidth]{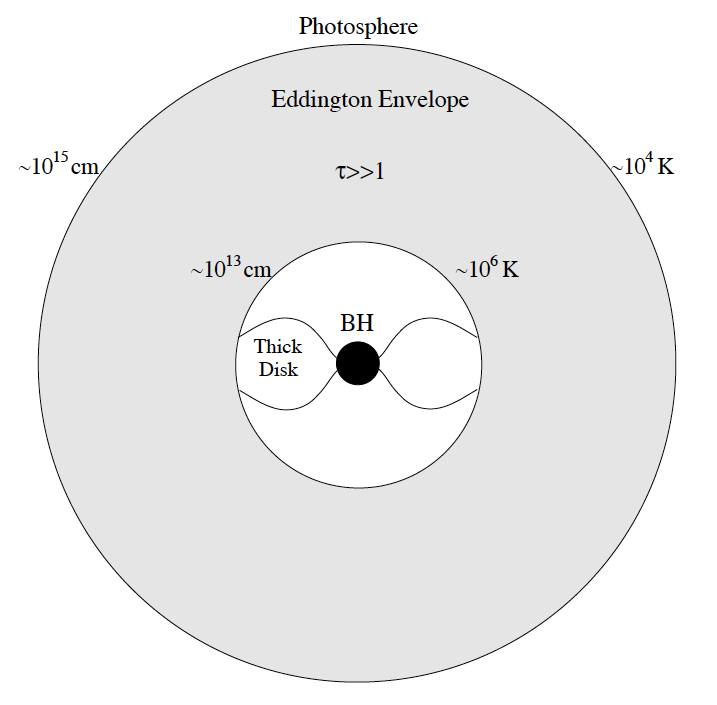}
\caption{Illustration of a steady, optically-thick, quasi-spherical large envelope surrounding a small accretion disc formed in TDEs. The disc X-ray/EUV emissions are reprocessed into UV/optical emissions when going through the envelope. Figure is from \citet{Loeb97}.}
\label{fig:envelope}
\end{center}
\end{figure}

Early studies \citep[e.g.,][]{Rees88} assume that stellar debris can circularize fast and form a compact disc with a size $R_c = 2~R_p$ from conservation of angular momentum, where $R_p$ is the pericenter radius of the original parabolic orbit of the star. Due to the short viscous timescale of the disc, the debris accretion rate resembles the fallback rate. Therefore, for the first few weeks to months, the disc in the super-Eddington accretion regime is geometrically and optically thick.
\citet{Ulmer99} calculates the effective temperature of this small, quasi-spherical disc by assuming the luminosity is limited by the Eddington luminosity $L_{\rm Edd}$ and the disc radius is comparable to the tidal radius $R_t$:
\begin{equation}
    T_{\rm eff} = \left( \frac{L_{\rm Edd}}{4\pi{R_t^2}\sigma} \right) \approx 5\times10^5~M_6^{-1/4} {\rm K},
    \label{eq:Teff}
\end{equation}
where $\sigma$ is the Stefan-Boltzmann constant and $M_6=M_{\rm BH}/(10^6 M_\odot)$.

This simple model predicts that TDEs should shine in X-ray/EUV and can therefore explain the observed X-ray strong TDEs (see the \xraychap~of this book). However, it cannot explain why many TDEs are only bright in optical/NUV bands but weak in X-rays (see the \optchap). It has been proposed by \citet{Loeb97} that since the stellar debris is only marginally bound, radiation produced from the viscous process in the inner accretion disc can easily disperse some gas into a quasi-spherical configuration at a large distance.
This large, steady, optically-thick envelope surrounding the small accretion disc, as illustrated in Fig. \ref{fig:envelope}, can reprocess the disc X-ray/EUV emissions into optical emissions.
The envelope is supported by radiation pressure and has a density profile of $\rho(r)\propto r^{-q}$ with $q=3$. It is assumed to be in a steady state exactly balanced by gravity and radiation pressure, which relies on a super-Eddington accretion rate with a radiative efficiency of $\gtrsim 10^{-3}$. The photosphere shines at the Eddington luminosity with a size of about $10^{15}$ cm and an effective temperature of around $10^4$ K.

\begin{figure}
\begin{center}
\includegraphics[width=0.8\textwidth]{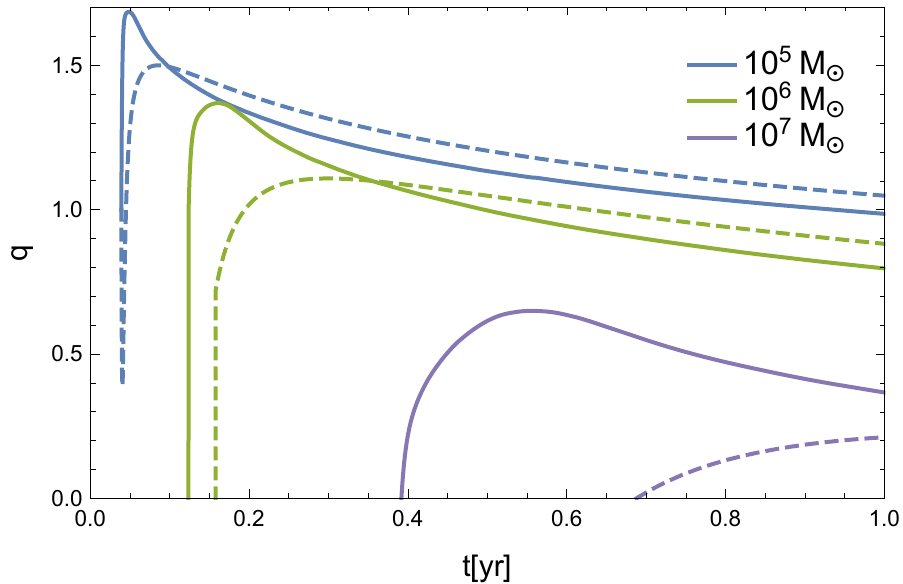}
\caption{Index $q$ of the zero-Bernoulli accretion (ZEBRA) flow \citep{Coughlin14} density profile ($\rho(r)\propto r^{-q}$) as a function of time for TDEs around black holes of different masses. The solid lines show results using numerically acquired fallback rate, and the dashed lines show results using the ``frozen-in''
approximation at the time of disruption. Figure is from \citet{Wu18}.}
\label{fig:ZEBRA}
\end{center}
\end{figure}

Later \citet{Coughlin14} argue that the angular momentum transport between the disc and envelope in the setting of \citet{Loeb97} should further shrink the disc and flatten its density distribution, making it hard to maintain the accretion exactly at the Eddington level. Instead, they propose a model in which the whole accretion disc is inflated by radiation energy into a quasi-spherical configuration with zero Bernoulli parameter (i.e., marginally bound) everywhere (these models are called ``Zero-Bernoulli accretion flows", or ZEBRAs). The disc rotates at a sub-Keplerian speed with a very small angular momentum consistent with the angular momentum of the returning stellar debris. The evolution of the accretion flow has been computed using the analytical debris fallback rate derived from the ``frozen-in'' approximation \citep{Lodato09} as well as using the numerically computed fallback rate \citep{Wu18}. As shown in Fig. \ref{fig:ZEBRA}, the density profile of the accretion flow $\rho(r)\propto~r^{-q}$ has an index $q$ ranging between 0.5 and 1.5 during most events. The photosphere radius grows with time to $R_{\rm ph}\sim {\rm few}\times10^{14}$ cm and the photosphere temperature drops by a few times to $T_{\rm ph} \sim {\rm few}\times10^4$ K after a few years. In this model, most of the energy produced from accretion is actually released in the form of a jet along polar directions confined by the narrow rotational funnel.

This type of spherical and steady envelope model has also been adopted by e.g., \citet{Roth16} to carry out detailed radiative transfer studies in TDEs. 

\subsubsection{Models of TDE discs with winds} \label{sec:wind}

A small fraction of the energy produced through accretion or other energetic processes (such as shock heating during disc formation) can in principle unbind a large fraction of returning debris and launch strong winds.
Wind-launching from accretion discs can be due to: 1) large radiation pressure in super-Eddington accretion discs or other radiative transfer processes in accretion discs \citep[e.g.,][]{Poutanen07, Proga00, Dotan11}; 2) hydromagnetic processes \citep[e.g.][]{Blandford82}; 3) in TDEs,  gas heating during the processes leading to disc formation such as collisions of debris steams \citep{Jiang16a}. 

Winds have been observed in several TDEs. For example, ultra-fast winds with $v\gtrsim 0.1c$ have been found through X-ray reflection signatures in Swift J1644 \citep{Kara16} and through X-ray absorption features or radio signals found in (soft) X-ray strong TDE ASASSN 14-li \citep{Alexander16, Kara17} and 3XMM J1500 \citep{Lin17}. Recently UV spectroscopic studies have also been fruitful in discovering winds with $v\approx0.05c$ from optical strong TDEs  \citep{Hung19}.

In the context of TDE discs, there are several studies modeling or simulating the winds launched during the super-Eddington accretion phase \citep{Strubbe09, Strubbe11, Lodato:2011a, Dai18, Curd19}. \citet{Strubbe09} consider a simple model in which the radiation energy responsible for wind-launching is generated when returning stellar debris continuously hits the already formed disc at a distance close to $R_c = 2R_p$. A constant fraction $f_{\rm out}\equiv \dot{M}_{\rm wind} / \dot{M}_{\rm fb}\approx0.1-0.3$ of the fallback debris is assumed to become a spherical wind, which is launched at $R_c$ with a velocity close to the escape velocity at $R_c$ and adiabatically expands. Inside $R_c$ a slim disc model similar to \citet{Abramowicz88} is adopted to describe the disc properties.  In particular, the disc temperature $T_{\rm d}$ is obtained as follows:
\begin{equation}
\sigma_{\rm SB}T_{\rm d}^4 = \frac{3GM_{\rm BH}\dot{M}f}{8\pi R^3}\left[\frac{1}{2}+\left(\frac{1}{4}+\left(\frac{\dot{M}}{\eta\dot{M}_{\rm Edd}}\right)^2\left(\frac{R_{\rm s}}{R}\right)^2\right)\right]^{-1},
	\label{eq:t}
\end{equation}
where $f=1-\sqrt{R_{\rm in}/{R}}$, $R_{\rm s}$ is the Schwarzschild radius and the accretion rate in the disc is given by $\dot{M}=(1-f_{\rm out})\dot{M}_{\rm fb}$.
In this model, strong optical emissions can be produced from the wind which cools down to few$\times10^4$ K after adiabatic expansion. The total emission is dominated by wind emission at earlier time (for $\sim10-100$ days depending on the exact stellar and black hole parameters and the wind launching radius) when the wind is optically thick. At later times, the wind becomes more dilute and the photosphere recedes, revealing more disc emissions.

\begin{figure}
\begin{center}
\includegraphics[width=0.8\textwidth]{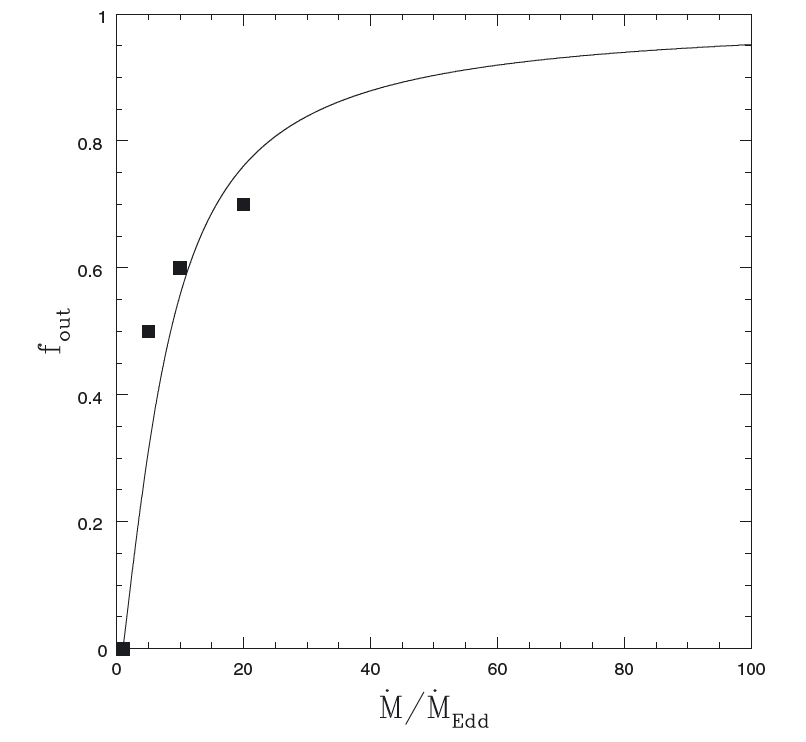}
\caption{The solid line shows the wind mass rate fraction $f_{\rm out}$ as a function of the fallback rate of stellar debris in unit of the Eddington accretion rate, which is used by \citet{Lodato:2011a} to model the wind launched in TDE super-Eddington accretion discs. As the Eddington ratio increases, more fallback debris material becomes wind instead of accreting onto the black hole. The squares show the numerical results of \citet{Dotan11}. Figure is from \citet{Lodato:2011a}.}
\label{fig:fout}
\end{center}
\end{figure}

\begin{figure}
\begin{center}
\includegraphics[width=0.8\textwidth]{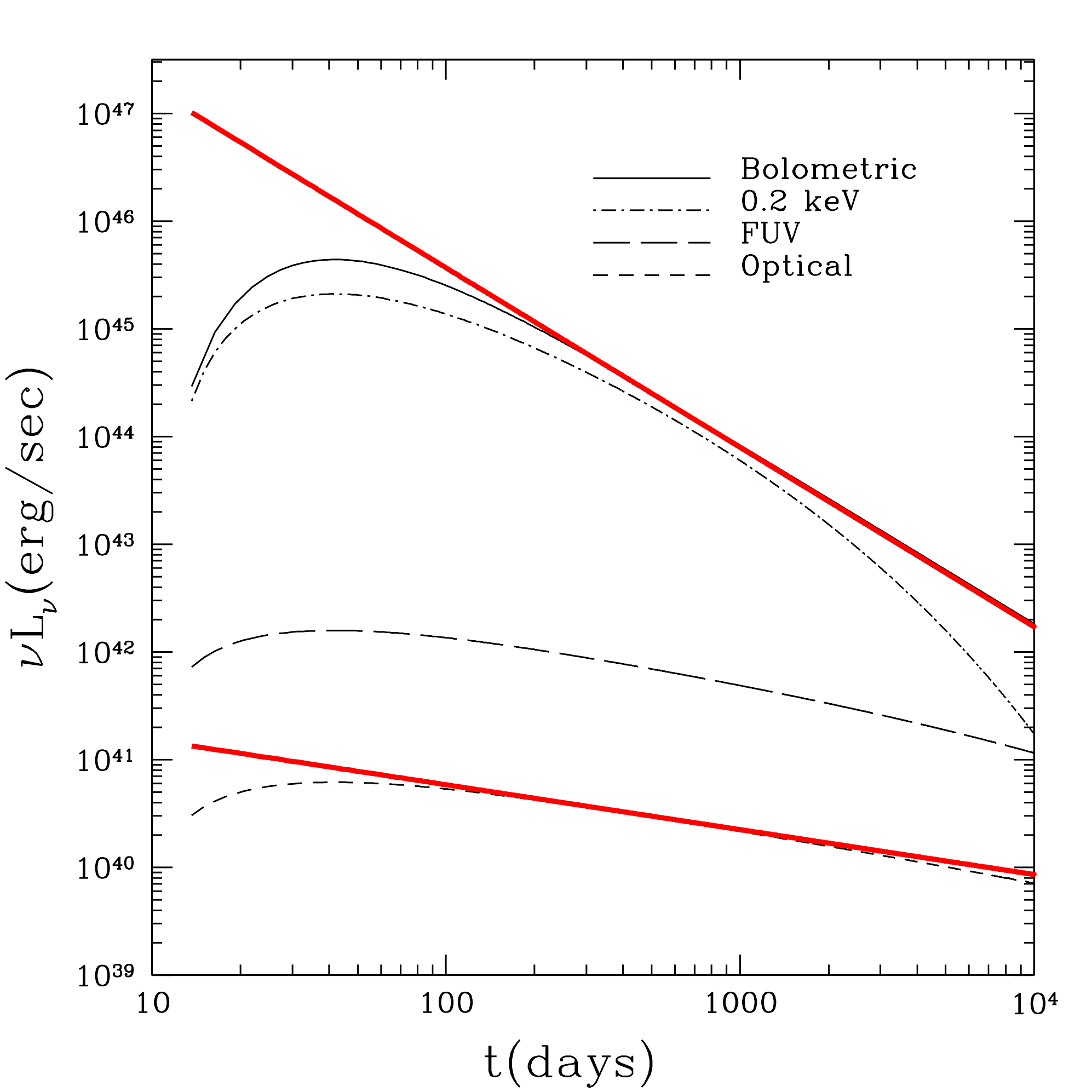}
\caption{Example lightcurves for the disc emission from the disruption of a solar type star by a $10^6M_{\odot}$ black hole at 
$\beta=1$. The fallback rate corresponds to a $\gamma=1.4$ polytropic model for the star \citep{Lodato09}. 
The solid line shows the bolometric luminosity, the dot-dashed line shows the luminosity at 0.2 keV, the long-dashed line shows the luminosity in the FUV, at $\nu = 1.9~10^{15}$ Hz, while the short-dashed line shows the luminosity in the optical at $\nu = 6.3~10^{14}$ Hz. The two red lines mark the simple power laws expected for the bolometric luminosity ($\propto t^{-5/3}$) and for the monochromatic luminosity in the optical/UV ($\propto t^{-5/12}$). Figure is from \citet{Lodato:2011a}.}
\label{fig:LR}
\end{center}
\end{figure}

\citet{Lodato:2011a} consider a more realistic model for the wind produced from the radiation-pressure supported disc formed in the super-Eddington phase, based on the simulation of \citet{Dotan11}. The wind mass loss fraction $f_{\rm out}$ is taken to be larger at earlier times when the debris supply rate is higher (Fig. \ref{fig:fout}). They also calculate the emission produced by the slim, compact disc, assuming that the accretion rate through the disc is a constant fraction of the fallback rate. They find that, even if the disc bolometric luminosity scales as the fallback rate, as $t^{-5/3}$, the lightcurve in specific bands scales differently with time, because the disc temperature also evolves as the accretion rate declines. In particular, at wavelengths (such as in the UV and optical) that sit on the Rayleigh-Jeans tail of the disc multi-color blackbody, the lightcurve is proportional to the temperature and therefore scales as $T_{\rm d}\propto L_{\rm bol}^{1/4}\propto t^{-5/12}$, declining much slower than the fallback rate. Conversely, for those wavelengths that are close to the peak of the blackbody at the beginning of the flare (i.e., in the soft X-rays, typically), the lightcurve  initially declines as $t^{-5/3}$ (see Fig. \ref{fig:LR}). In general, the lightcurve is expected to evolve from a $t^{-5/12}$ decline, to a $t^{-5/3}$ and eventually to an exponential decline with time.  For high enough photon energies (e.g., in the X-rays), the first phase could be missing, and the ligthcurve would start directly from a close to $t^{-5/3}$ decline. If one also adds the wind emission in addition to the disc emission, the lightcurve gets further complicated and depends strongly on the system parameters, with the apparent lightcurve evolving coincidentially as $t^{-5/3}$ in particular regions of the parameter space. 

These two models discussed previously assume that a sizable fraction of the fallback debris can make its way to the inner disc and super-Eddington accretion (onto the black hole) still happens. Instead, \citet{MillerCole15} and \citet{Metzger16} propose models in which most of the debris supply is launched as winds faraway from the black hole, either due to large opacities in the outer disc or due to shocks happening during the disc formation process. Under such scenarios, only a small fraction ($\lesssim 1\%$) of debris materials is accreted so the accretion rate is always sub-Eddington. The low level of accretion, together with the very mass-loaded wind functioning as a  thick reprocessing envelope, can also reduce temperature of the TDE emission to few$\times10^4$ K so that UV/optical emissions dominate.

Furthermore, numerical simulations of TDE discs with super-Eddington accretion rates \citep{Dai18, Curd19} also show that strong winds can be launched in this phase, and we will discuss these numerical work in the next section.

\subsubsection{Simulated super-Eddington accretion flow and inclination dependence}

Realistically, we do not expect accretion flows to be spherically symmetric.  While the studies mentioned in Section \ref{sec:steadymodel} and \ref{sec:wind} mostly adopt 1D disc or envelope models when carrying out the semi-analytical calculations, they also suggest that the structure and emission of TDE accretion flow should have a dependence on the observer's viewing-angle with respect to the disc orientation. The details, however, can only be calculated using numerical simulations due to the complexity of the problem.

Codes that can be used to calculate the dynamics of super-Eddington accretion discs and capture its basic radiative transfer processes have only been developed in the recent decade \citep{Ohsuga05, Ohsuga11, Jiang14, Sadowski13, McKinney14}. (See the \simchap~of this book for details.) These novel simulations reveal that super-Eddington accretion discs always launch strong, optically-thick winds due to their large radiation pressure. The winds obscure the discs from a large solid angle but has an inclination-dependent profile. Close to the pole, wind is dilute and ultra-fast speed ($v\gtrsim 0.1c$). Close to the disc, wind is dense and moves slower.
Analytical models of super-Eddington accretion discs, such as the slim disc model discussed previously, however, usually do no predict super-Eddington discs launch winds.

\begin{figure}
\begin{center}
\includegraphics[width=1.0\textwidth]{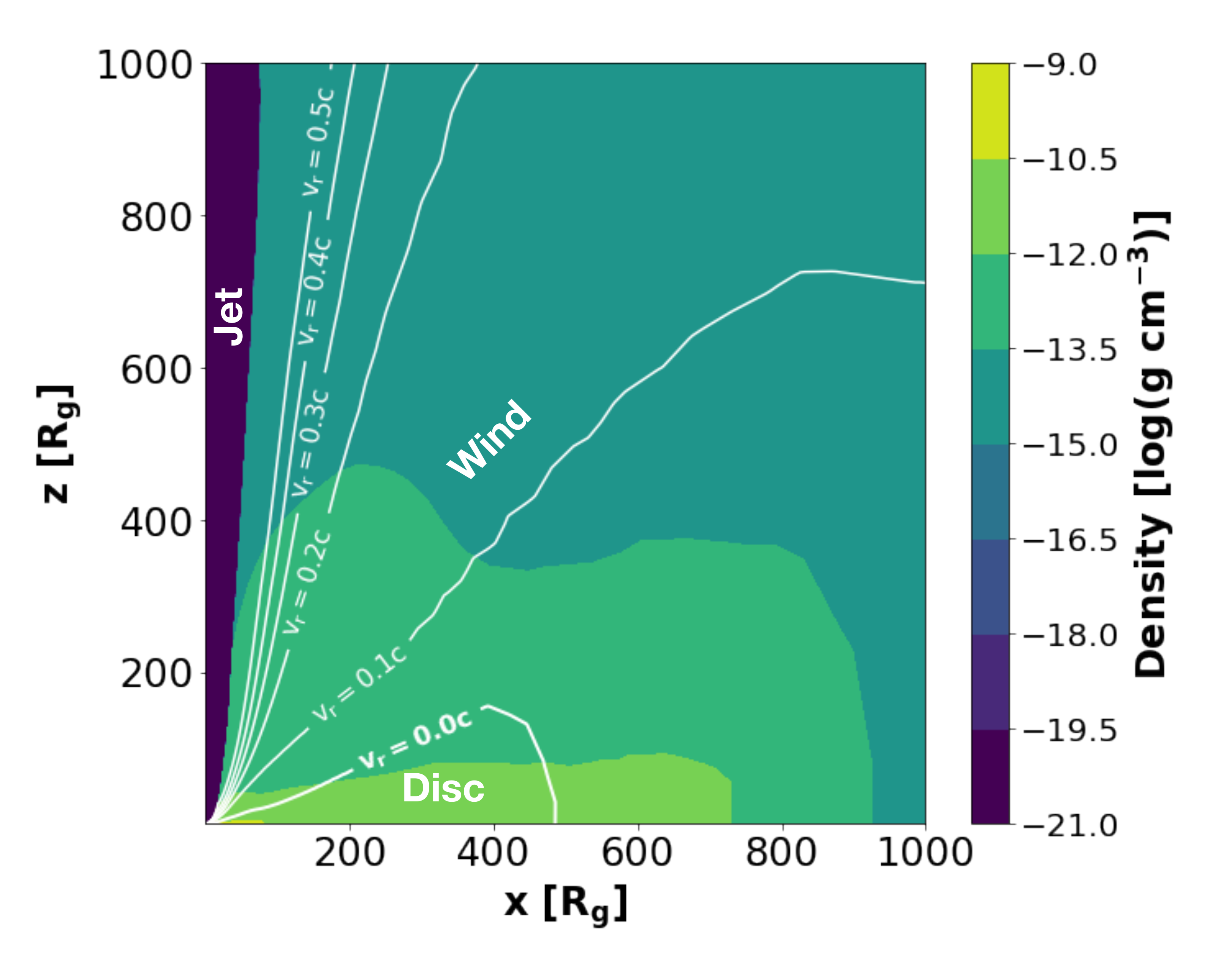}
\caption{Disc, wind and jet regions of the  super-Eddington accretion flow simulated in \citet{Dai18}. The density of the disc and winds are
shown by the background color. The white lines are contours of constant radial velocity in the wind region. Closer to the pole, the wind is more dilute and moves faster. The jet region (where magnetic pressure dominates the gas pressure) is marked using a dark blue colour. Figure is adapted from \citet{Thomsen19}.}
\label{fig:supereddstruture}
\end{center}
\end{figure}

In order to check the  viewing-angle dependence of TDE super-Eddington discs, \citet{Dai18} have carried out a 3D general-relativistic radiation magneto-hydrodynamic (GRRMHD) simulation using a code called \textsc{HARMRAD} \citep{McKinney14, McKinney15}. For the initial condition, they use a small, Keplerian, geometrically thick disc with mass and angular momentum about one tenth of the original stellar mass and angular momentum. The simulation has been evolved to achieve an inflow equilibrium of about $100R_g$ (which is about ten times larger compared to previous super-Eddington disc simulations), in order to ensure resolving the photosphere from most of the inclination angles. The structure of the accretion disc and wind can be seen in Fig. \ref{fig:supereddstruture}. The accretion disc is geometrically and optically thick. The wind, launched by a combination of radiation and magnetic pressure, is also optically thick from most inclinations except from at angles close to the pole. A jet is also launched by the Blandford-Znajek mechanism due to the set up of the simulation with a large black hole spin and sufficient poloidal magnetic field threading the disc.

Reprocessing of the inner disc X-ray emissions happens in the wind and outer disc. At the same time, the wind also expands and cools down. Therefore, one should expect the wind emission to shine at lower energy bands due to the both adiabatic expansion and reprocessing as discussed in e.g., \citet{Strubbe09, Roth16, Metzger16}. The simulated disc profile is then post-processed using a Monte-Carlo radiative transfer code \textsc{SEDONA} \citep{Kasen06, Roth16} to obtain the spectra from different inclinations.
It is found that UV/optical emission dominates when viewed along the disc direction, and X-ray emission dominates from directions closer to the pole. Interestingly, it is also found that most of the radiation energy will escape in the extreme ultraviolet (EUV) waveband which is heavily absorbed from extragalactic sources, and this might explain the ``missing-energy'' issue of the observed TDEs \citep{Lu18}. 
Based on the results, \citet{Dai18} propose a unified model for the observed UV/optical and X-ray TDEs based on the viewing-angle dependent structure of the accretion flow.

\citet{Curd19} further simulate TDE super-Eddington discs with the GRRMHD code \textsc{KORAL} \citep{Sadowski13}, using the ZEBRA model as disc initial conditions. While the first-order disc and wind structures are similar to those of \citet{Dai18}, since the initial large-scale structure of the accretion flow used in their simulations is more isotropic, the escaped emissions from different inclinations demonstrate less variance. 

Both \citet{Dai18} and \citet{Curd19} have started the simulations with already circularized geometrically-thick small discs aligned with the black hole spin. Therefore, they cannot address the concerns related to the disc formation, such as whether TDE discs can indeed circularize promptly. To our knowledge, \citet{Sadowski16tde} has carried out the only MHD simulation of TDEs from the disc formation phase to the accretion phase. Due to the constraint on computational time, the parameters chosen (i.e., an initial eccentric stellar orbit with a penetration parameter of 10) may only represent a small fraction of TDEs.  Interestingly, it is found that while the disc can circularize fast, turbulence inherited from the violent debris stream self-interaction persists in the accretion flow, sustained either by convection or by the fall-back of gas.  This hydrodynamic turbulence dominates over the magnetorotational instability (MRI) in the differentially rotating flow and contributes significantly to the angular momentum transfer.

\subsection{Late-time TDE disc models with disc spreading}
\label{sec:thindisc}

\begin{figure}
\begin{center}
\includegraphics[width=0.9\textwidth]{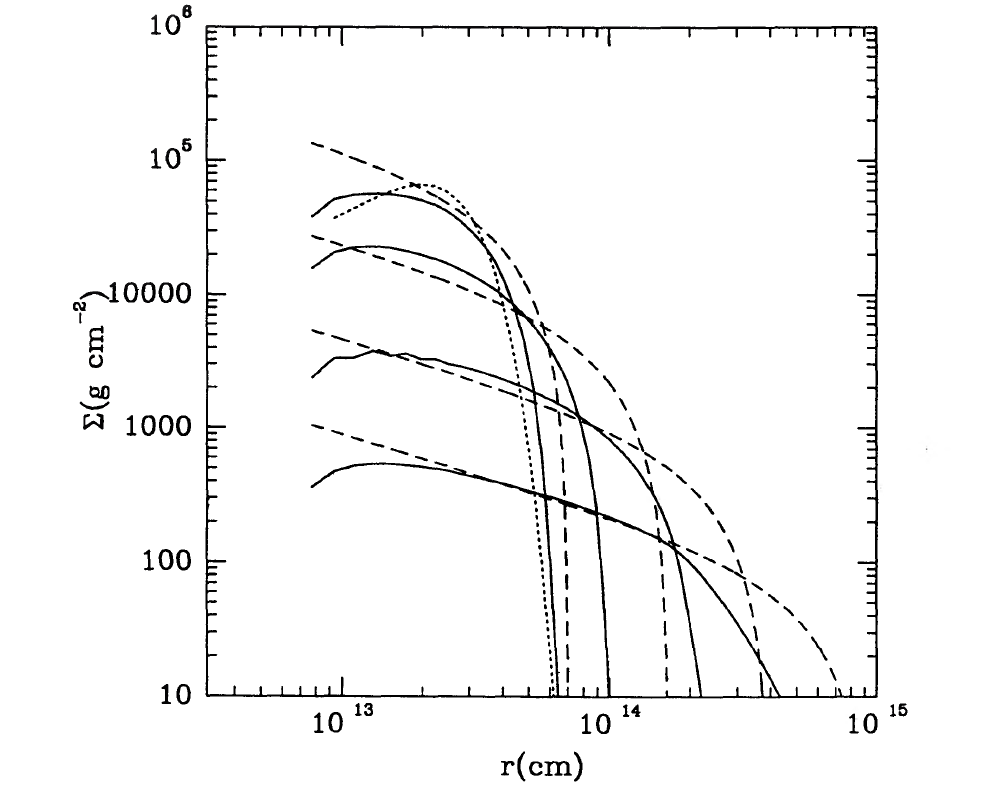}
\caption{Evolution of the surface density in the disc model by \citet{Cannizzo90}. The ring-structured disc has a mass of $0.5 M_\odot$ and viscosity parameter $\alpha=1$. The disc continues to spread out and surface density drops with the disc spreading. Curves show the surface density profile at the times of 0, 3, 30, 300,and 3000 yrs. Figure is from \citet{Cannizzo90}.}
\label{fig:disclate}
\end{center}
\end{figure}

At late time in a TDE, as the accretion level gradually drops from super-Eddington to sub-Eddington regime, the disc thickness decreases and eventually should become a geometrically thin.
The disc viscous or inflow timescale greatly increases with the reduced disc height. If the disc viscous timescale exceeds the debris fallback timescale, the draining rate of the disc through accretion will become lower than the mass supply rate of the returning debris, so material will accumulate in the disc. 

\citet{Cannizzo90} have calculated the evolution of a disc which initially has a ring structure with maximum surface density at $R \sim R_t$ and a total mass equal to a sizable fraction of the stellar mass. The disc is modeled using a standard $\alpha$-disc prescription with a viscosity parameter $\alpha = 0.01 -1$. As disc spreads both outwards and inwards due to gas pressure, its surface density decreases as shown in Fig. \ref{fig:disclate} and the disc temperate drops. While the exact accretion rate depends on the initial disc mass as well as the value of the disc viscosity and opacity, at later times the accretion rate  adopts the same decline pattern with $\dot{M}\propto t^{-1.2}$, which is flatter than the canonical debris fallback rate. This model has been further revised by \citet{vanVelzen19} and \citet{Mummery20} to explain the late-time light curves of TDEs observed in the UV or X-ray bands \citep[e.g.,][]{Gezari17, Holoien18}.

It has been a long-standing unsolved problem whether run-away thermal instability happens in radiation pressure-dominated thin accretion discs \citep{Shakura73}. \citet{Shen:2014b} investigate this topic in the context of TDEs discs and predict that thermal instability, if it happens, can lead to cycles of bursts of accretion. So far observed TDE flares have not demonstrated such behaviors. This perhaps suggests that thin discs formed in TDEs can stand against thermal instability either due to the support from high iron opacity \citep{Jiang16b} or large-scale vertical magnetic field \citep{Liska19a}, or due to the suppression of thermal instability from large-scale toroidal magnetic field \citep{Begelman07}.

\subsection{Eccentric accretion flow formed in TDEs}
\label{sec:eccentric}

\begin{figure}
\begin{center}
\includegraphics[width=0.8\textwidth]{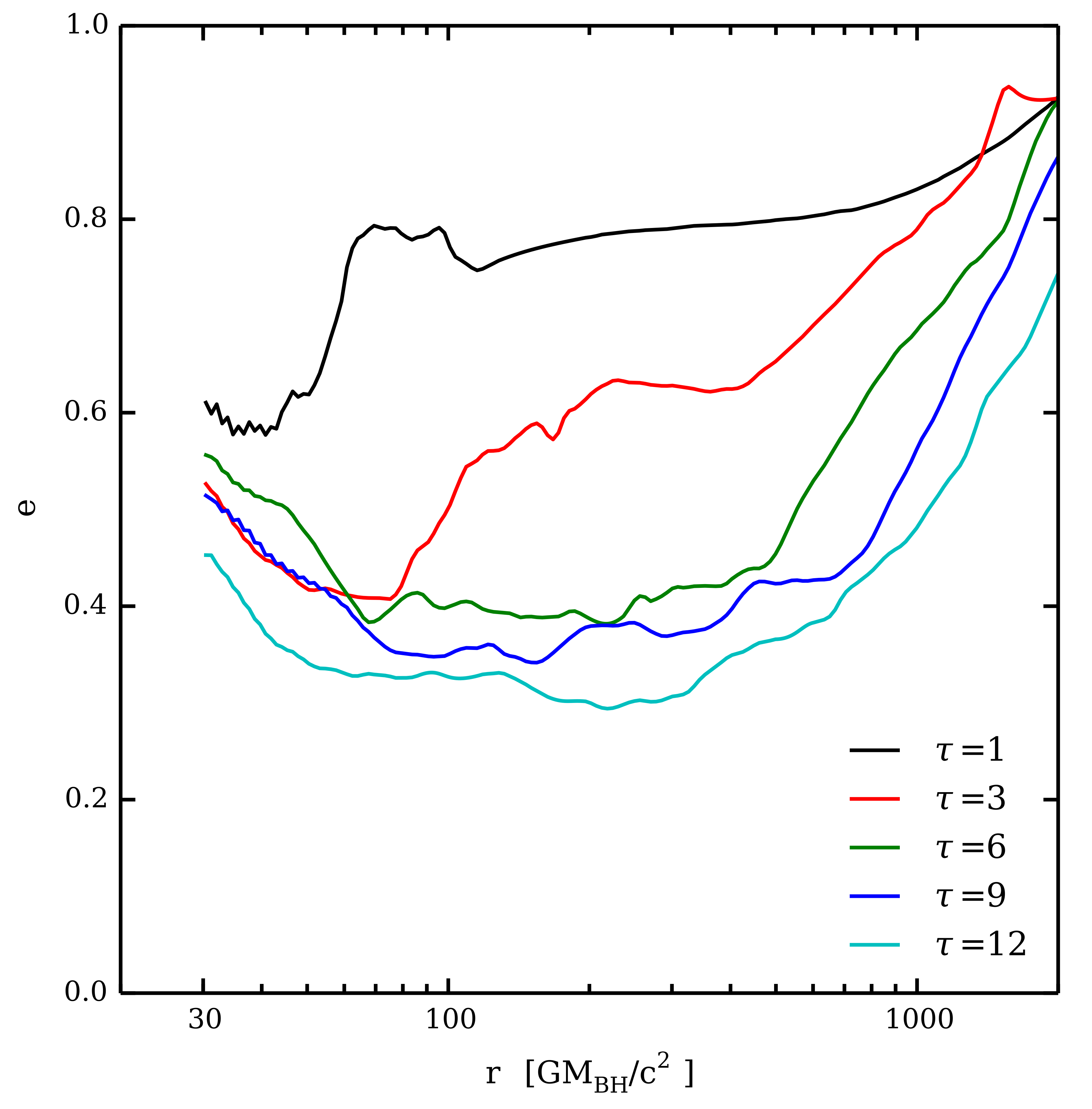}
\caption{Radial profile of eccentricity from the simulated accretion flow in \citet{Shiokawa15}.  The time evolution of mass-weighted mean eccentricity is given in units of $\tau$, the orbital timescale of the most tightly bound matter. Note that the inner radius in this model is truncated at $30 R_g$. Figure is from \citet{Shiokawa15}.}
\label{fig:eccentricdisc}
\end{center}
\end{figure}

As current simulations show that nozzle shock happening at pericenter or debris stream self-crossing due to GR apisdal precession may not dissipate orbital energy fast enough,
it is suggested that the just-formed disc can have an eccentric configuration. For example, for the stellar and black hole parameters in \citet{Shiokawa15}, the simulated accretion flow after several fall-back timescales is moderately eccentric.  From the paper, Fig. \ref{fig:eccentricdisc} shows the time evolution of the radial profile of mass-weighted mean eccentricity, with the eccentricity settling to $e\sim0.3-0.4$, extending from the inner-radius of the simulation ($\sim 30 R_g$) to the minimum semi-major axis of the most bound debris ($\sim 500 R_g$).  Note that the tidal radius for this simulation is $\sim 100 R_g$, and the general relativistic hydrodynamics simulation does not include magnetic fields.  These results, large disc and slow time scale for circularization, are contrary to the canonical model, where it is assumed that the accretion flow circularizes on the order of a few times the orbital timescale of the most tightly bound matter. The distinctive structure and time scale of the accretion flow imply that optical emission in TDEs could possibly be produced by mechanisms different from accretion onto the black hole.

It is also a topic under investigation whether eccentric accretion flow around black holes can generally circularize on timescale faster than the viscous timescale. 
In analytic treatments to the dynamics of eccentric discs, \citet{Syer92} suggest that an initially eccentric accretion flow can maintain its configuration or even become more elliptical by applying Gauss perturbation equations upon a $\alpha$-viscosity disc model. \citet{Ogilvie2014} and \citet{Barker2014} show that the eccentric disc evolution can depend on extreme, non-linear instabilities, providing a useful tool for understanding the leading mechanisms especially with follow-up numerical investigations.  Based on this framework, \citet{Chan2018} show, with a toy model, how MRI growth occurs close to pericenter and declines with increasing eccentricity of the debris orbits. One remark is that the picture can become more complicated if one also considers general relativistic procession of the debris stream orbits and large-scale magnetic fields in the disc, which is beyond the capacity of analytical treatments.

In light of this, several studies have turned towards emission produced during the evolution of an elliptical accretion flow.  Such mechanisms are perhaps at the location of stream collision, at apocenter $a_{\rm min}$ when the GR apsidal precession is weak \citep{Piran15,Krolik16}. There are also many open questions in the evolution of eccentric accretion discs in the context of TDEs. For example, \citet{Svirski2017} address the issue of angular momentum loss by proposing a direct plunge of debris onto the black hole due to MHD stresses. We refer the readers to the \flowchap~in this book for more details.

Certain observational features of TDEs, especially double-peaked or flat-topped broad H$\alpha$ emission lines seen in a few candidates \citep[e.g.,][]{Holoien2019}, have motivated studies adopting thin elliptical disc models to explain the line profiles  \citep{Liu2017,Cao2018}.  For example, \citet{Eracleos1995} give expressions for the total line flux from the disc, which depend on a relativistic Doppler factor for photons in an elliptical structure, namely the eccentricity and inclination angle.  Alternatively, such line profiles may also be modeled using circular discs with optically-thick winds \citep{Roth2018,Hung19}.

\subsection{Angular momentum transport in TDE discs} \label{sec:newLtransport}

The issue of angular momentum transport in accretion discs has always been a significant source of controversy and investigations. For years, a phenomenological description based on a viscous disc model has been used, following the pioneering work of \citet{Shakura73}, until it was recognized that turbulence induced by the magneto-rotational instability (MRI) is the most likely origin for the required stress within the disc (see, e.g. \citealt{Balbus98} and references therein). Are discs originating from TDEs consistent with the established models for longer lived discs around Active Galactic Nuclei (AGN) and galactic black hole binaries? The question arises because the initial magnetic field of the star is expected to be relatively low, and the process of disruption itself is not expected to provide a significant enhancement in the magnetic field strength \citep{Bonnerot17,Guillochon17}. 

Roughly speaking, one would imagine that, in order for the MRI to be established, the magnetic field should be strong enough that the Alfven speed is comparable to the gas sound speed. A full understanding of how MRI functions and how magnetic fields grow via MRI requires resolving the microphysical dissipation processes at small scales in the plasma. Global disc simulations have so far demonstrated that MRI can still kick in at significantly small magnetic field strengths, corresponding to a plasma $\beta$ value  (the ratio of thermal to magnetic pressure) of the order of $\sim 100-1000$ \citep{Bugli18,Zhu18}. If we apply this value for the conditions of typical of TDE discs, this would correspond to $B\sim 10^{3-4}$ G, much larger than the expected $1$~G field strength corresponding to the unmperturbed star. 
Recently, \citet{Nealon18} have demonstrated how the torus resulting from a TDE can be unstable to the Papaloizou-Pringle instability \citep{Papaloizou84}, that occurs for thick torii which are supported mainly by pressure against gravity. Such instability can provide a significant means of transport of angular momentum over the first $\sim 20$ orbits of the torus, before the MRI eventually develops, with the effective viscosity in the range of $\alpha\sim 0.03-0.05$. In addition, it could also provide a time-varying quadrupole moment that can give rise to potentially detectable gravitational wave emission \citep{Toscani19}.

\section{Jets Launching in TDEs: How And When}
\label{sec:tdejet}

So far three TDEs have been observed to produce powerful jets: Swift J1644+57 \citep{Bloom11, Burrows11, Zauderer:2011}, Swift J2058+05 \citep{Cenko12} and Swift J1112-82 \citep{Brown15}. Their beamed hard X-rays/soft $\gamma$-rays and the accompanied strong radio emissions have been interpreted to originate from relativistic, collimated jets formed in TDEs. This gives the most direct evidence that transient accretion discs form in TDEs. As we have discussed in Section 2, TDE discs transition from the super-Eddington accretion regime to the sub-Eddington thin disc regime and then eventually to the ADAF/RIAF regime, in a timescale much faster than typical AGN state-changing timescales. Therefore, TDEs can also be used to study jet physics in all three accretion regimes. In this section, we mainly focus on the jet-launching physics, and leave the jet emission physics to be discussed in details in the \emischap, \xraychap~and \radiochap. 

The best studied jet launching mechanism is the Blandford-Znajek (hereafter BZ) mechanism, in which relativistic jet can be powered by extracting the BH rotation energy via ordered magnetic field threading the ergosphere of a spinning black hole \citep{Blandford77}. In the standard BZ approximation, the BH energy extraction rate scales as $a^2 ~\Phi^2_B$ ($a$ is the black hole spin parameter and $\Phi_B$ is the large-scale magnetic flux around the black hole). This has been demonstrated in GRMHD simulations \citep[e.g.,][]{McKinney05, Tchekhovskoy11, McKinney12}. Under certain conditions the jet power can increase even faster than $a^2$ for large valules of $a$ \citep{Tchekhovskoy12}.

The accretion flow around an astrophysical BH plays a crucial role in transporting magnetic flux to its vicinity, amplifying magnetic flux via MRI or dynamo process, and confining magnetic flux from leaking \citep[e.g.,][]{Pessah05, Beckwith09}.
Since it is easier for magnetic flux to accumulate in geometrically thick discs \citep{Lubow94}, powerful, continuous jets are expected to be launched in the super-Eddington regime or the ADAF/RIAF regime. Furthermore, the thick disc geometry can help jet collimation. 

\begin{figure}
\begin{center}
\includegraphics[width=1.0\textwidth]{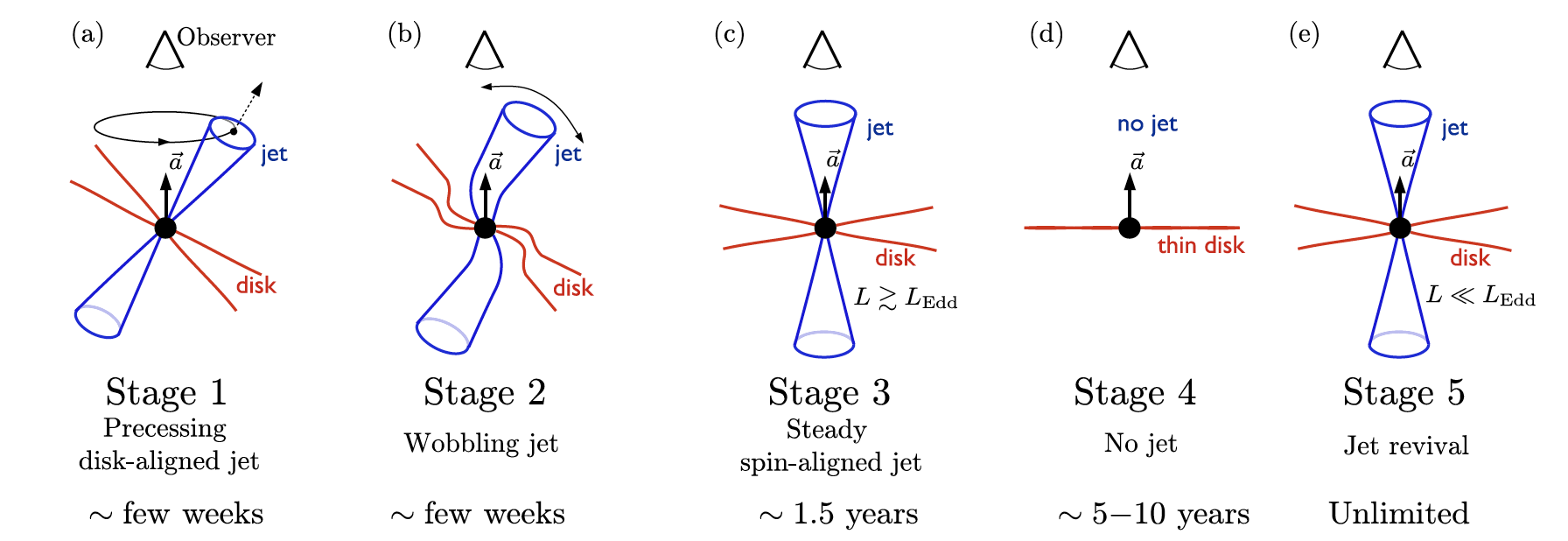}
\caption{Evolution of the jet and disc in Swift J1644 proposed by \citet{Tchekhovskoy14}. A misaligned jet initially forms in the super-Eddington accretion phase, which precesses and wobbles around the black hole spin axis. Within weeks the jet orientation is aligned with black hole spin axis due to the magneto-spin alignment mechanism. After about one year, the disc becomes sub-Eddington, so the jet is quenched. As the accretion accretion continues to drop, the jet is predicted to revive when the disc eventually transitions to the ADAF/RIAF accretion regime. Figure is from \citet{Tchekhovskoy14}.}
\label{fig:tdejet}
\end{center}
\end{figure}

The change of accretion regime has been used to explain the the X-ray light curve of the Swift J1644 event \citep{Zauderer13, Tchekhovskoy14}: Initially when the debris disc accretion rate was super-Eddington and the disc was geometrically thick, a relativistic, collimated jet was launched, producing strong X-ray and radio emissions. During this phase the jet power and X-ray flux roughly traced the debris fallback rate following the $t^{-5/3}$ pattern since about 10 days after the trigger. Then the X-ray emission rapidly decreased by two orders of magnitude after about 500 days, by which time the fallback rate was expected to drop to around or below the Eddington level. As a consequence, the disc morphology changed and the jet quenched. This evolution, together with the initial disc-jet alignment process which will be discussed in Section \ref{sec:discalign}, is illustrated in Fig. \ref{fig:tdejet}.

When are the most powerful jets produced? For any accretion disc there exists an upper limit of the magnetic flux that the disc can confine near the black hole. When the magnetic flux exceeds this limit, the magnetic pressure becomes stronger than the disc gravity and pushes the accretion flow away from the black hole. The type of accretion flow under the strong magnetic-field limit is called the magnetically-arrested disc (MAD) \citep{Narayan03}. One can define a dimensionless magnetic flux parameter:
\begin{equation}
    \Upsilon \equiv \frac{0.7~\Phi}{\sqrt{4\pi R^2_g \dot{M}c^2}},
     \label{eq:Upsilon}
\end{equation} 
where $\Phi$ is the flux of poloidal magnetic flux threading the disc in Gaussian units and $\dot{M}$ is the accretion rate. The disc is in the MAD state when $\Upsilon\gg1$.

GRMHD simulations show that the most powerful jets with large Lorentz factor are launched from MAD discs, which can achieve a jet efficiency $\eta_{\rm jet} \equiv L_{\rm jet}/\dot{M}_{\rm BH}c^2 \approx 10-100\%$  \citep{Hawley06, Tchekhovskoy11, Tchekhovskoy12, McKinney12}. On the other hand, weakly magnetized discs in the standard and normal evolution (SANE) regime \citep{Narayan12} can only produce low-power jets or non-relativistic outflows in the polar region \citep{DeVilliers05, Moscibrodzka16}. 
Recent GR-radiation-MHD simulations of super-Eddington (TDE) discs also show consistent results \citep{McKinney15, Sadowski16, Dai18, Curd19}. 

If the Swift J1644 jet is launched from a MAD disc by the BZ mechanism, then the magnetic flux threading the BH needed to explain the peak X-ray luminosity is about $10^{30}~{\rm G~cm^2}\times (M_{\rm BH}/10^6M_\odot)$ \citep{Bloom11, Kelley14}, which is 3--5 orders of magnitude larger than the original magnetic flux possessed by a main-sequence star. This therefore has triggered a series of investigations on how magnetic flux can be amplified during the disruption, disc forming or accretion processes in TDEs. Simulations studying the magnetic field amplification during the disruption process \citep{Guillochon17,Bonnerot17} show that although the dynamo process in surviving cores from partial disruption events might be important, the amplification in returning debris in a total disruption event is not significant enough to supply the flux needed for the Swift J1644 jet. From the prospect of magnetic-field growth via MRI in accretion discs, \citet{Krolik11} propose that Swift J1644 might have been produced by an intermediate mass black hole (IMBH) disrupting a white dwarf. In this scenario the disc orbital timescale is shorter than the case of a SMBH disrupting a main-sequence star, so the magnetic field can quickly grow via turbulence to reach saturation. (However, \citet{Kara16} later constrains the SMBH mass in Swift J1644 to be a few$\times10^6~M_\odot$ from X-ray reverberation studies, which favors the disruption of a main-sequence star instead.) \citet{Kelley14} propose another model in which extra magnetic fields can be collected by the returning debris when it sweeps through a ``fossil'' disc left from previous AGN activities. When the stellar orbit and the fossil disc aligns, the magnetic flux that a quiescent fossil disc can provide to the debris is enough to power the Swift J1644 jet.

Alternatively, it has been suggested that in the TDE super-Eddington accretion phase a jet can be driven by the large radiation pressure and collimated by the gemetrically thick accretion flow \citep{Coughlin14, Sadowski15}. This jet production mechanism, however, has not been studied in details. More studies on this topic are needed to conclude whether relativistic jets can be produced purely due to radiation pressure.

\section{Precession And Alignment of  TDE Discs And Jets}\label{sec:discalign}

One peculiar feature of TDEs is that in general we expect the orbit of the incoming star (and thus the angular momentum of the debris) to be randomly oriented with respect to the spin of the central supermassive black hole, with the latter determined  by  the merger  history  of the host galaxy and the accretion history of the black hole.  Since the stellar pericenter is relatively close to the gravitational radius of the hole, we expect the debris to be significantly affected by relativistic effects in the form of Lense-Thirring precession. Such effects might delay the formation of a disc, by avoiding stream collisions due to a randomization of the orbital plane \citep{Dai13,Guillochon15, Hayasaki16}, or might speed-up circularization, by producing a thicker torus \citep{Gafton19}. Either way, after the disc is formed, it is expected that the initial orientation of the disc will be misaligned with the black hole spin due to conservation of angular momentum. Also, misaligned discs may also launch jets, and the jet orientation can be affected by both the disc angular momentum and the black hole angular momentum. In general, since such misalignment are specific to TDEs and might strongly affect the evolution of the system, several studies have investigated the dynamics of misaligned discs and jets in TDEs \citep{Stone12, Lei13, Tchekhovskoy14, Franchini16,Xiang16,Ivanov18,Zanazzi19}.

\subsection {Expected precession regimes in TDE discs}

The main physical effect at work in misaligned discs around a spinning black hole is Lense-Thirring precession. The precession torque at radius $R$ is given by (see, for example, \citealt{Franchini16} and references therein):
\begin{equation}
\mathbf{T}=-\Sigma R^{2}\Omega\left(\frac{\Omega_{z}^{2}-\Omega^{2}}{\Omega^{2}}\right)\frac{\Omega}{2}\mathbf{e_{z}\times l}\label{eq:torque-vect},
\end{equation}
where $\Sigma$ is the disc surface density, ${\bf l}$ is the unit vector along the local disc angular momentum direction, ${\bf e}_z$ is the direction of the black hole spin, and $\Omega$ and $\Omega_z$ are the angular and vertical frequencies, respectively, given by:
\begin{equation}
\Omega=\frac{c^{3}}{GM}\frac{1}{r^{3/2}+a}\label{eq:omega};
\end{equation}
\begin{equation}
\left(\frac{\Omega^{2}-\Omega_{z}^{2}}{\Omega^{2}}\right)=4ar^{-3/2}-3a^{2}r^{-2}\label{eq:omega_z},
\end{equation}
where $M$ is the black hole mass, $r$ is the disc radius in units of the gravitational radius of the hole, and $a$ is the black hole spin parameter.

The response of the disc to external misaligned torques has been studied extensively over the years. The pioneering work of \citet{Papaloizou83} and of \citet{Papaloizou95} have shown that the disc responds to such perturbations either in a diffusive way (when the ratio of turbulent stresses to pressure $\alpha$ is larger than the disc aspect ratio $H/R$) or in a wave-like way (in the opposite regime, when $\alpha\lesssim H/R$). A full non-linear theory of diffusive warp propagation has been provided by \citet{Ogilvie99}. On the other hand, a non-linear theory of bending waves has not been developed, and only a linear \citep{Lubow2000} and quasi-linear theory \citep{Ogilvie06} exist. Non-linearities might prevent the wave-like regime from happening altogether \citep{Hawley18}. For thin and viscous discs, or for discs that evolve diffusively, the inner disc aligns rapidly to the black hole spin (on the warp diffusion timescale), while the outer disc retains its original direction and slowly aligns at longer times. This is known as the Bardeen-Petterson effects \citep{Bardeen75}.

In the case of TDE discs, we expect initially the disc to be thick (large $H/R$) due to the high accretion level and with a relatively low value of internal ``viscous'' stresses, as the magneto-rotational instability is building up (low $\alpha$), and should therefore be characterized by bending wave propagation. At later times, however, the disc becomes thinner and magnetic stresses build up, so a transition to the diffusive regime is expected to occur.

\subsection{Global disc and jet precession in the thick disc regime}

At early times in TDEs, regular global precession of the disc might occur, as shown in the GRMHD simulations of thick torii \citep{Fragile05, Fragile07, Liska18}. For TDEs, this was initially suggested by \citet{Stone12} and later confirmed with time-dependent calculations by \citet{Franchini16}. The global precessional frequency can be computed easily as a weighted average of the Lense-Thirring frequency $\Omega_{\rm LT}$ \citep{Stone12,Franchini16}:
\begin{equation}
\Omega_{\mathrm{p}}=\frac{\int_{R_{\mathrm{in}}}^{R_{\mathrm{out}}}\Omega_{\mathrm{LT}}(R)L(R)2\pi R\, dR}{\int_{R_{\mathrm{in}}}^{R_{\mathrm{out}}}L(R)2\pi R\, dR},\label{eq:freqP}
\end{equation}
where $L(R)=\Omega R^2\Sigma$. Equation (\ref{eq:freqP}) is applicable when the disc is narrow (as expected for TDE) and the warp is mild (Note that a mild warp does not imply a mild misalignment). The Lense-Thirring frequency is easily derived from Eqs. (\ref{eq:torque-vect}) and (\ref{eq:omega_z}):
\begin{equation}
\Omega_{\mathrm{LT}}(r)=\frac{1}{2}\frac{c^{3}}{GM}\frac{1}{r^{3/2}+a}(4ar^{-3/2}-3a^{2}r^{-2}).\label{eq:local-prec}
\end{equation}

\begin{figure}
\begin{center}
\includegraphics[width=0.8\textwidth]{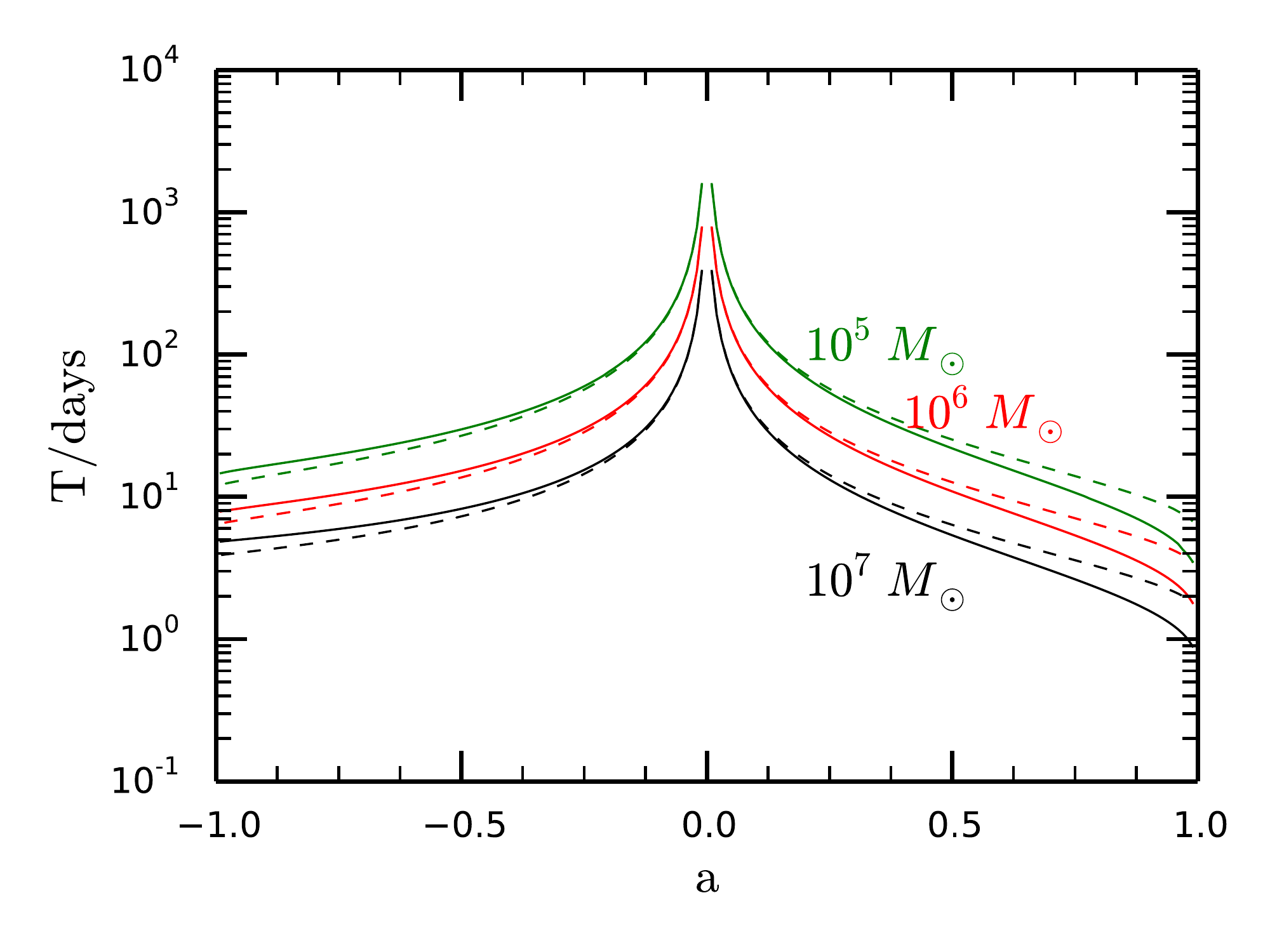}
\caption{Precession period expected from a TDE disc for three different choices of SMBH mass (green: $10^5M_{\odot}$, red: $10^6M_{\odot}$, black: $10^7M_{\odot}$. Solid line refer to the full relativistic calculation, while dashed line adopt a first order approximation to the Lense-Thirring frequency. These model refer to disc that extend out to $2R_{\rm t}$, for $\beta=1$. Figure is from \citet{Franchini16}}
\label{fig:franchini}
\end{center}
\end{figure}

The resulting precession periods are in the range of a few to several days, depending on black hole mass and spin, and become longer for lower spins and lower black hole masses (see Fig. \ref{fig:franchini}, taken from \citealt{Franchini16}). However, it should be noted that such periods strongly depend on where the disc lies, given the steep scaling of the Lense-Thirring rate with radius. If, for some reason, the disc lies much closer to the hole, the precession rate could be much higher. This will also occur if the disc is torn into smaller rings, in cases where the misalignment is very large \citep{Nixon12}.

A few recent studies have explored the effect of the torques induced by infalling material from the debris stream onto the disc. It is found that the effect of such torques is to complicate the disc dynamics, leading to irregular precessional motion of the disc \citep{Xiang16,Ivanov18}, possibly giving rise to quasi-periodic, rather than period precession.

Recent GRMHD simulations have shown that tilted discs can also launch relativistic jets \citep{McKinney13, Liska18}. When the disc is not highly magnetized, the jet has been shown to initially follow the disc angular momentum orientation on large scales and precess together with the disc, although the tilt angle between the black hole spin and the disc/jet decreases over longer timescales. (See section \ref{sec:magnetospin} for the alternative case of highly magnetized discs.) 

\subsection {Viscous vs cooling driven disc alignment}

Over long timescales, independently on the original misalignment, the disc is expected to settle down on the equatorial plane of the black hole due to dissipative effects \citep{Sorathia13a,Sorathia13b},  although some non dissipative relaxation processes have been sometimes suggested for stellar systems \citep{Merritt12}, the applicability of which to TDE discs has not been shown yet. \citet{Stone12} argue that this would have naturally happened due to the fact that, as the accretion rate through the disc decreases (for example, following the fallback rate, as $t^{-5/3}$), the disc would become thinner and gradually move to the Bardeen-Petterson regime, at which stage rapid alignment would follow. However, even in the wave-like propagation phase, alignment is expected to happen as a consequence of viscous dissipation.
\citet{Franchini16} evaluate specifically the relevant alignment process in the presence of both cooling driven disc thinning and of viscous effects in the disc. The two possible alignment processes have a distinctly different dependence on the viscosity parameter $\alpha$, whereby cooling induced alignment scales as $\alpha^{-3/5}$, while viscous induced alignment scales as $\alpha^{-1}$. \citet{Franchini16} explore a wide range of parameters and conclude that for most cases, viscously induced alignment occurs more quickly that cooling induced alignment. The only cases for which the disc aligns due to the Bardeen-Petterson effect are for large black hole mass, low spins and low values of $\alpha$. 

The alignment time is of the order of tens to hundreds of days for large black hole spin and grow to a few years for very low spins.

\subsection {Magneto-spin alignment of strongly magnetized discs and jets} \label{sec:magnetospin}

When large-scale electromagnetic (EM) field exists around the black hole, EM force can also affect the dynamics of jets and discs and provides another torque to affect their orientations. 
In the case of a highly magnetized disc, the strong coupling between the black hole spin and its magnetosphere can make not only the relativistic jet but also the inner part of a thick accretion disc align with the black hole spin axis. 

One can compare the EM force torque and the torque induced by the Lense-Thirring effect to see which one dominates the disc dynamics. With simple assumptions such as that the magnetic field, when being pushed against by a disc, bends on the same scale of the disc radius $r$, \citet{McKinney13} calculate that: 
\begin{equation}
    \frac{\tau_{\rm EM}}{\tau_{LT}} \approx \frac{1}{64~a}~\Upsilon^2~\frac{r}{R_g}~\alpha_{\rm eff}~\big(\frac{H}{R}\big)^2.   
\end{equation}
Here $a$ is the dimensionless black hole spin parameter, $\Upsilon$ is the dimensionless magnetic flux given by Eqn. \ref{eq:Upsilon}, $H/R$ is the disc aspect ratio, and $\alpha_{\rm eff}\equiv$ $v_r/(v_\phi~(H/R)^2)$ is the effective viscosity parameter.
Therefore, the EM torque becomes more dominant when the disc is more magnetized, or when the disc is thicker.
3D GRMHD simulations and calculations \citep{McKinney13, Polko17, Liska18} have further demonstrated that for the highly magnetized thick discs (such as MAD discs) with a large range of tilt angles, the EM torque is dominant and strong enough to reorient the inner disc and inner jet to align with the BH spin axis.

Tidal disruption events likely produce thick accretion discs due to the high accretion level. These discs might also be strongly magnetized, especially when relativistic jets are produced which require large-scale magnetic fluxes to power (see Section \ref{sec:tdejet}).  Therefore, EM force could play an important role in aligning the discs and jets with black hole spins in TDEs such as Swift J1644.

\subsection{Warping and tearing of thin accretion discs}

In this regime which is typically assumed to happen at late times in TDEs, we expect the inner part of a thin disc not to precess regularly but to simply align with the black hole spin axis due to the Bardeen-Petterson effect. This has been demonstrated in Smoothed Particle Hydrodynamics simulations by \citet{Nelson00, Lodato07, Lodato10} as well as in GRMHD simulations by \citet{Liska19a, Liska19b}. As the disc is fed by debris on fallback timescales, the outer part of the disc may not have sufficient time to be aligned and its orientation could be still determined by the angular momentum of the falling back debris. The disc can become warped or even torn apart \citep{Nixon12, Liska19b}. \citet{Zanazzi19} compute eigenmodes for the disc warp excited by Lense-Thirring precession and apply it to disc produced in TDEs. They compute global precession frequencies finding a broad agreement with those predicted by \citet{Franchini16}, with precession periods ranging from days to weeks depending on the disc parameters. 

\subsection{Possible cases of misaligned discs and jets observed in TDEs}

\subsubsection{QPO in ASASSN 14-li}

From the observational side, ASASSN14-li interestingly shows a quasi-periodic oscillation in the soft x-ray band ($0.3-1$ keV) with a periodicity of only 131 seconds with $3-4 \sigma$ significance \citep{Pasham19}. It has been argued that this QPO is possibly associated with the global Lense-Thirring precession of a misaligned disc, though quantitative calculations pose several challenges. From Eqn.[\ref{eq:local-prec}] such a high precession rate would require an extremely high spin for the black hole even if using the LT precession rate at the black hole ISCO radius. This case is further puzzling since the signal has been observed to be stable for roughly 450 days, which is much longer than the expected alignment process for such a high spin.

\subsubsection{Behavior of the relativistic jet in Swift J1644}

The X-ray level of Swift J1644 shows large-amplitude variability in the first few weeks \citep{Bloom11}, which has been interpreted to be the result of jet wobbling \citep{Tchekhovskoy14}. Then the jet and the highly magnetized disc have been aligned by the magneto-spin alignment mechanism as discussed in Section \ref{sec:magnetospin}, as shown in Fig. \ref{fig:tdejet}. This model is further supported by the lack of clear evidence of jet precession during the main light curve decay phase lasting for about one year, although there is a report of a $\sim4\sigma$ detection of a 200 second QPO in the $2-10$ keV X-ray band with an unknown nature \citep{Reis12}.

\section{Summary And Future Studies}
\label{sec:summary}

The studies of accretion discs, winds and jets in TDEs discussed in this review add an important component to the studies of general accretion and jet physics around black holes. The extreme level of debris accretion and the fast transition across different accretion regimes offer us unique opportunities of studying physics that do not in happen other accretion phenomena such as AGNs or X-ray binaries.

The masses and spins of the SMBHs contain valuable information about their growth as well as the assembling and merger history of the galaxies since the early universe. 
TDEs are important probes to the demographics of SMBHs, since they can illuminate dormant SMBHs, which represent the majority of the SMBH population at low redshifts, with bright flares lasting for years. Also, while AGN observations tend to select luminous, massive targets, TDEs allow us to study less massive galaxies and their SMBHs, since stars approaching black holes with $M_{\rm}\gtrsim 10^8 M_\odot$ will typically plunge into the event horizon before being tidally disrupted (unless the SMBH is spinning very rapidly, see \citealt{Kesden12}).

A good understanding of the accretion and jet physics in TDEs will help us independently constrain SMBH parameters from TDE observables. For example, the power of a magnetically driven jet strongly correlates with the value of the black hole spin. Also, while the debris mass fallback rate has a dependence on the SMBH mass (e.g., $\dot{M}_{\rm fb, peak}\propto M^{-1/2}_{\rm BH}$, as in \citealt {Guillochon13}) and therefore the TDE light curve can in principle be used to constrain the SMBH mass \citep{Mockler19}, the luminosity (in monochromatic bands) may not exactly scale with the fallback rate, since the overall radiative effciency or the SED of the emission can change with accretion rate \citep{Lodato:2011a, Piran15b}, letting alone the uncertainties and delays related to the disc formation process.

As of now most of the studies of TDE accretion flow have focused on the ideal case of circular and aligned discs. They have shown that winds and relativistic jets can be launched at early times when the accretion is in the super-Eddington regime and the disc geometry is thick. These studies also offer explanations for the basic emission properties of TDEs, such as how X-ray emission produced close to the black hole can be reprocessed into UV/optical emissions in optically-thick winds or outer accretion flow. Studies of TDE discs at late times in the sub-Eddignton phase show that such unobscured thin discs should mostly produce X-ray emissions. They have long viscous timescales so they can keep accumulating mass and significantly spread outwards on debris fallback timescales. 
Recently, more explorations have been made into less understood regimes such as misaligned discs/jets or eccentric accretion discs. A lot of the progress as discussed in this chapter has been made possible largely thanks to various novel numerical simulations carried out recently to investigate accretion flow and jets which are too complex to be studied analytically.  

At the end of this review, we pose a few open questions to motivate future studies: 
\begin{itemize}
\item[$\ast$] Recently more TDEs have been followed up for several years past the peak of the flare. This offers a good opportunity for further studying thin accretion discs and especially the transition between the super-Eddington regime and the thin disc regime. For example, do TDE emissions at late times share similar properties as AGN emissions which are assumed to be produced from thin accretion discs? 
\item[$\ast$] If, in a TDE, the disc is initially highly eccentric, how does such a disc evolve?  Can it maintain an eccentric configuration for a long time or circularize quickly? How does MRI work in such discs? Can eccentric discs also generate winds and jets?
\item[$\ast$] The basic emission properties of TDEs and some of their spectroscopic features can be qualitatively explained using current accretion and emission models. For example, from the observed Bowen fluorescent lines in several optically strong TDEs, we know from first order calculations that there exists an optically-thick reprocessing layer that converts X-ray/EUV emissions into optical emissions \citep{Leloudas19}. Also the asymmetric, broad emission line profiles in some TDEs \citep{Hung19, Holoien2019} indicate that these discs either produce winds or have an eccentric geometry. In the future more quantitative studies along these lines will be very useful for studying the properties of the accretion flow. 
\item[$\ast$] How can we apply TDE accretion and jet physics to study other accretion phenomena, such as AGNs, Galactic X-ray binaries, ultra-luminous X-ray sources and gamma-ray bursts? And vice versa? 
\item[$\ast$] The emission from a typical X-ray strong AGN always has a power-law hard X-ray component, which is believed to be produced by the Compotonization of the cold disc photons by a hot ``corona'' above the disc. However, (non-jetted) X-ray strong TDEs only produce soft thermal emissions. What makes coronae disappear in TDE discs?
\item[$\ast$] If most TDE discs are surrounded by large, optically-thick reprocessing layers (in the form of steady envelopes or winds), then it may not be straightforward to directly probe the 
space-time very close to black holes with TDE observables. Perhaps X-ray strong TDEs and jetted TDEs give us the best hope for constraining black hole spins.
\end{itemize}

In the future, detailed models and simulations of disc formation, accretion and emission processes will help us better answer these questions in our quest of probing the SMBH demographics, black hole accretion/jet physics and general relativity with potential samples of thousands of tidal disruption events.

\begin{acknowledgements}
The authors thank the International Space Science Institute (ISSI) for their support and hospitality and the review organisers for their leadership in coordinating these reviews. The authors also thank the Yukawa Institute for Theoretical Physics (YITP) at Kyoto University for hosting the International Molecule-type Workshop ``Tidal Disruption Events: General Relativistic Transients'' (YITP-T-19-07), where the authors held useful discussions to complete this work. The authors are grateful to the referees for providing constructive suggestions and comments. JLD is  supported  by  the  GRF  grant  from the Hong Kong government under HKU 27305119. The work of RMC is funded by the Advanced Simulation Computing Physics and Engineering Program under the auspices of Los Alamos National Laboratory, operated by Triad National Security, LLC, for the National Nuclear Security Administration of U.S. Department of Energy (Contract  No. 89233218CNA000001). 

\end{acknowledgements}


\bibliographystyle{aps-nameyear}

\bibliography{main}

\end{document}